\PassOptionsToPackage{table,dvipsnames}{xcolor}
\documentclass{article}

\usepackage{microtype}
\usepackage{graphicx}
\usepackage{booktabs} 

\usepackage{hyperref}

\usepackage[accepted]{icml2025}

\usepackage{amsmath}
\usepackage{amssymb}
\usepackage{mathtools}
\usepackage{amsthm}

\usepackage[capitalize,noabbrev]{cleveref}

\theoremstyle{plain}
\newtheorem{theorem}{Theorem}[section]

\theoremstyle{definition}
\newtheorem{definition}[theorem]{Definition}

\theoremstyle{remark}

\usepackage[textsize=tiny]{todonotes}

\usepackage{pifont}
\usepackage{multirow}
\usepackage{makecell}
\usepackage[caption=false]{subfig}
\usepackage{tcolorbox}
\newtcolorbox{dialogbox}{
  arc=4mm,
  colback=blue!3,
  colframe=black,
  rounded corners,
  boxrule=0.5pt,
  fonttitle=\bfseries,
  coltitle=black,
}
\usepackage{listings}
\definecolor{codegreen}{rgb}{0,0.6,0}
\definecolor{codegray}{rgb}{0.5,0.5,0.5}
\definecolor{codepurple}{rgb}{0.58,0,0.82}
\definecolor{backcolour}{rgb}{0.95,0.95,0.92}

\lstdefinestyle{mystyle}{
    backgroundcolor=\color{backcolour},   
    commentstyle=\color{codegreen},
    keywordstyle=\color{magenta},
    numberstyle=\tiny\color{codegray},
    stringstyle=\color{codepurple},
    basicstyle=\ttfamily\footnotesize,
    breakatwhitespace=false,         
    breaklines=true,                 
    captionpos=b,                    
    keepspaces=true,                 
    numbers=left,                    
    numbersep=5pt,                  
    showspaces=false,                
    showstringspaces=false,
    showtabs=false,                  
    tabsize=2
}

\usepackage{xspace}

\newcommand{\figref}[1]{\figurename~\ref{#1}}

\newcommand{\tabref}[1]{\tablename~\ref{#1}}
\newcommand{\algref}[1]{Algorithm~\ref{#1}}
\newcommand{\secref}[1]{Section~\ref{#1}}

\newcommand{\bheading}[1]{{\vspace{0.3\baselineskip}\noindent{\textbf{#1}}}}

\newcommand{\eg}{\textit{e.g.}\xspace}

\definecolor{light-gray}{gray}{0.9}

\icmltitlerunning{EvoVerilog: Large Langugage Model Assisted Evolution of Verilog Code}

\begin{document}

\twocolumn[
    \icmltitle{EvoVerilog: Large Langugage Model Assisted Evolution of Verilog Code}



    \icmlsetsymbol{equal}{*}

    \begin{icmlauthorlist}
        \icmlauthor{Ping Guo}{equal,cityu}
        \icmlauthor{Yiting Wang}{equal,maryland}
        \icmlauthor{Wanghao Ye}{maryland}
        \icmlauthor{Yexiao He}{maryland}
        \icmlauthor{Ziyao Wang}{maryland}\\
        \icmlauthor{Xiaopeng Dai}{amazon}
        \icmlauthor{Ang Li}{maryland}
        \icmlauthor{Qingfu Zhang}{cityu}
    \end{icmlauthorlist}

    \icmlaffiliation{cityu}{Department of Computer Science, City University of Hong Kong, Hong Kong}
    \icmlaffiliation{maryland}{Department of Electrical Engineering, University of Maryland, Maryland, United States}
    \icmlaffiliation{amazon}{Amazon, Seattle, United States}
    \icmlcorrespondingauthor{Ang Li}{angliece@umd.edu}
    \icmlcorrespondingauthor{Qingfu Zhang}{qingfu.zhang@cityu.edu.hk}

    \icmlkeywords{Machine Learning, ICML}

    \vskip 0.3in
]



\printAffiliationsAndNotice{\icmlEqualContribution} 

\begin{abstract}
    Large Language Models (LLMs) have demonstrated great potential in automating the generation of Verilog hardware description language code for hardware design. This automation is critical to reducing human effort in the complex and error-prone process of hardware design.
    However, existing approaches predominantly rely on human intervention and fine-tuning using curated datasets, limiting their scalability in automated design workflows.
    Although recent iterative search techniques have emerged, they often fail to explore diverse design solutions and may underperform simpler approaches such as repeated prompting.
    To address these limitations, we introduce EvoVerilog, a novel framework that combines the reasoning capabilities of LLMs with evolutionary algorithms to automatically generate and refine Verilog code.
    EvoVerilog utilizes a multiobjective, population-based search strategy to explore a wide range of design possibilities without requiring human intervention.
    Extensive experiments demonstrate that EvoVerilog achieves state-of-the-art performance,  with pass@10 scores of 89.1 and 80.2 on the VerilogEval-Machine and VerilogEval-Human benchmarks, respectively. Furthermore, the framework showcases its ability to explore diverse designs by simultaneously generating a variety of functional Verilog code while optimizing resource utilization.
\end{abstract}
\section{Introduction}
The emergence of Large Language Models (LLMs) has significantly advanced the automation of complex tasks across various domains, demonstrating remarkable capabilities in natural language understanding~\cite{hendrycks:2021:measuring}, mathematical problem solving~\cite{yu:2024:metamath}, and code generation~\cite{chen:2021:evaluating,liu:2024:systematic}.
In particular, LLMs have shown exceptional potential in code generation, enabling developers to automate the creation of code snippets, debug existing implementations, and even contribute to algorithm discovery~\cite{romera:2024:mathematical,liu:2024:eoh,guo:2024:coevo}.
These advances in structured text processing have facilitated innovative applications in Electronic Design Automation (EDA), a field critical to the design of modern integrated circuits (ICs)~\cite{zhong:2023:llm4eda,wu:2024:chateda,pearce:2020:dave}.

\begin{figure}[t]
    \centering
    \includegraphics[width=0.46\textwidth]{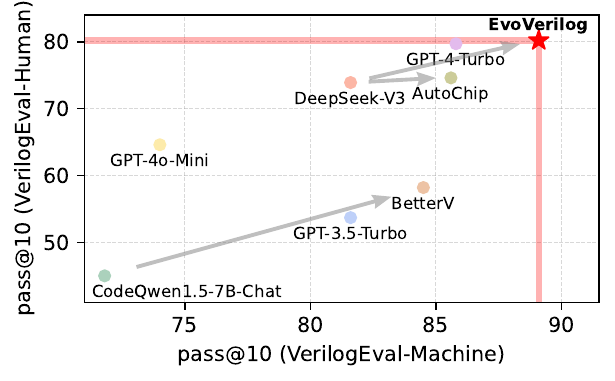}
    \caption{Comparison of pass@10 rates across different methods on the VerilogEval benchmark. EvoVerilog demonstrates superior performance, highlighting its effectiveness in automated Verilog code generation.\label{fig:method_comp}}
\end{figure}

LLMs have made substantial contributions to EDA algorithms and tools by improving language-related tasks in IC design,
particularly in generating hardware description language (HDL) code such as Verilog and VHDL~\cite{liu:2023:chipnemo}.
The growing complexity of ICs, driven by Moore's law, has significantly increased the time and expertise required for manual HDL coding.
Consequently, LLM-based automation has emerged as an indispensable solution to improve EDA productivity~\cite{zhong:2023:llm4eda}.

Early efforts in LLM-based Verilog code generation have primarily explored two approaches: \emph{1)} human interaction and \emph{2)} fine-tuned generation.
Initially, the focus was on the use of general-purpose LLMs, trained in vast Internet datasets, to assist human designers~\cite{blocklove:2023:chipchat,chang:2023:chipgpt}.
While pioneering, these approaches were limited by their reliance on human intervention.
Subsequently, researchers investigated fine-tuning LLMs using domain-specific data and techniques to improve their performance~\cite{liu:2024:rtlcoder,pei:2024:betterv}.
However, these methods face challenges of scalability and generalizability due to their high demand for training data and computational resources.
Furthermore, they primarily emphasize generating correct code in a single attempt,which diverges from real-world development practices that often involve iterative refinement.
As a result, these approaches fail to fully harness the reasoning capabilities of LLMs.

Recent advances have highlighted the potential of automated Verilog code design through iterative search~\cite{thakur:2023:autochip,tsai:2024:rtlfixer}, a paradigm that is closely aligned with real-world hardware development practices.
This approach utilizes feedback mechanisms to iteratively refine designs, thereby improving the quality of generated code.
However, existing iterative search approaches suffer from two key limitations: \emph{1)} the search process is often confined to a fixed initial solution, which restricts exploration of the design space and frequently often results in suboptimal outcomes; \emph{2)} current approaches lack support to address multi-objective design requirements, a critical aspect of hardware design where balancing resource efficiency with functional correctness is essential. These limitations underscore the necessity for a more efficient framework that facilitates iterative design refinement while simultaneously addressing multiple design objectives.

To address these limitations, we propose EvoVerilog, a novel framework that integrates the reasoning capabilities of LLMs with the search capabilities of evolutionary algorithms (EAs) for automated Verilog code generation.
The framework comprises two key components: 1) an idea tree-based generation process that enhances solution diversity, and 2) an EA-based mechanism that promotes solution convergence.
By combining the idea-tree generation process with the population-based search of EAs, the framework overcomes the first limitation by enabling broader exploration of the design space.
Additionally, we introduce a nondominated sorting scheme to further enhance solution diversity and address the second limitation by supporting multiobjective optimization.
Our framework achieves state-of-the-art performance on the widely adopted VerilogEval benchmark, as demonstrated by the comparison of pass@10 rates against existing methods in \figref{fig:method_comp}.

Our contribution can be summarized as follows:

\begin{itemize}
    \item \textbf{EvoVerilog Framework:} We introduce EvoVerilog, the first framework to combine LLMs with evolutionary algorithms for fully automated generation of Verilog code, eliminating the need for human intervention.
    \item \textbf{Idea-Tree-Based Search:} A novel idea tree generation process enhances the exploration of the design space, ensuring broader and more diverse solutions compared to fixed initialization approaches.
    \item \textbf{Multi-Objective Optimization:} Our framework incorporates a non-dominated sorting scheme to balance competing design objectives, such as functional correctness and resource efficiency, enabling diverse and high-quality solutions.
    \item \textbf{State-of-the-Art Performance:} EvoVerilog achieves pass@10 scores of 89.1 and 80.2 on the VerilogEval-Machine and VerilogEval-Human benchmarks, outperforming existing methods and setting a new standard for automated Verilog generation. Moreover, EvoVerilog can generate Verilog code that is both functionally correct and resource efficient, demonstrating its practical utility in real-world hardware design scenarios.
\end{itemize}
\section{Background}
\begin{table*}[t]
    \centering
    \caption{Comparative analysis of LLM-based methods for Verilog code generation, categorized by human interaction, fine-tuned generation, and iterative search approaches. It outlines the methods, their features, and the test datasets used for evaluation. Specifically, for the features of these methods (\textbf{M1-4}), \textcolor{green}{\ding{51}} indicates the presence of the feature, while \textcolor{red}{\ding{55}} indicates the absence.}
    \vspace{0.25\baselineskip}
    \label{tab:methods}
    \resizebox{0.98\textwidth}{!}{
        \begin{tabular}{l|l|cccc|l}
            \toprule
            \textbf{Category}                                           & \textbf{Method}                                             & \textbf{M1}                  & \textbf{M2}                  & \textbf{M3}                  & \textbf{M4}                  & \textbf{Test Dataset}                                                               \\
            \midrule
            \multirow[c]{2}{*}{\textbf{Human Interaction}}              & ChipChat~\cite{blocklove:2023:chipchat}                     & \textcolor{red}{\ding{55}}   & \textcolor{green}{\ding{51}} & \textcolor{green}{\ding{51}} & \textcolor{green}{\ding{51}} & Several small-scale circuits                                                        \\
            \cline{2-7}
                                                                        & ChipGPT~\cite{chang:2023:chipgpt}                           & \textcolor{red}{\ding{55}}   & \textcolor{green}{\ding{51}} & \textcolor{green}{\ding{51}} & \textcolor{red}{\ding{55}}   & Several small-scale circuits (for a large-scale circuit)                            \\
            \midrule
            \multirow[c]{5}{*}{\textbf{Fine-tuned Generation}}          & ChipNeMo~\cite{liu:2023:chipnemo}                           & \textcolor{red}{\ding{55}}   & \textcolor{red}{\ding{55}}   & \textcolor{red}{\ding{55}}   & \textcolor{red}{\ding{55}}   & Customized datasets + VerilogEval~\cite{liu:2023:verilogeval}                       \\
            \cline{2-7}
                                                                        & VeriGen~\cite{thakur:2024:verigen}                          & \textcolor{red}{\ding{55}}   & \textcolor{red}{\ding{55}}   & \textcolor{red}{\ding{55}}   & \textcolor{green}{\ding{51}} & 181 from HDLBits~\cite{wong:2024:hdlbits} + VerilogEval~\cite{liu:2023:verilogeval} \\
            \cline{2-7}
                                                                        & VerilogEval~\cite{liu:2023:verilogeval}                     & \textcolor{red}{\ding{55}}   & \textcolor{red}{\ding{55}}   & \textcolor{red}{\ding{55}}   & \textcolor{red}{\ding{55}}   & VerilogEval~\cite{liu:2023:verilogeval}                                             \\
            \cline{2-7}
                                                                        & RTLCoder~\cite{liu:2024:rtlcoder}                           & \textcolor{red}{\ding{55}}   & \textcolor{red}{\ding{55}}   & \textcolor{red}{\ding{55}}   & \textcolor{green}{\ding{51}} & VerilogEval~\cite{liu:2023:verilogeval} + RTLLM~\cite{yao:2024:rtllm}               \\
            \cline{2-7}
                                                                        & BetterV~\cite{pei:2024:betterv}                             & \textcolor{red}{\ding{55}}   & \textcolor{red}{\ding{55}}   & \textcolor{red}{\ding{55}}   & \textcolor{red}{\ding{55}}   & VerilogEval~\cite{liu:2023:verilogeval}                                             \\
            \midrule
            \multirow[c]{3}{*}{\textbf{Iterative Search}}               & AutoChip~\cite{thakur:2023:autochip}                        & \textcolor{green}{\ding{51}} & \textcolor{red}{\ding{55}}   & \textcolor{green}{\ding{51}} & \textcolor{green}{\ding{51}} & VerilogEval~\cite{liu:2023:verilogeval}                                             \\
            \cline{2-7}
                                                                        & RTLFixer~\cite{tsai:2024:rtlfixer}                          & \textcolor{green}{\ding{51}} & \textcolor{red}{\ding{55}}   & \textcolor{green}{\ding{51}} & \textcolor{green}{\ding{51}} & VerilogEval~\cite{liu:2023:verilogeval}                                             \\
            \cline{2-7}
                                                                        & \textbf{EvoVerilog (Ours)}                                  & \textcolor{green}{\ding{51}} & \textcolor{green}{\ding{51}} & \textcolor{green}{\ding{51}} & \textcolor{green}{\ding{51}} & VerilogEval~\cite{liu:2023:verilogeval}                                             \\
            \bottomrule
            \multicolumn{2}{l}{\textbf{M1}: Automatic Error Correction} & \multicolumn{5}{l}{\textbf{M2}: Multiple Design Objectives}                                                                                                                                                                                                                   \\
            \multicolumn{2}{l}{\textbf{M3}: Training-Free}              & \multicolumn{5}{l}{\textbf{M4}: Open Source }
        \end{tabular}
    }
\end{table*}
\subsection{LLMs for Verilog Code Generation}
Since early 2023, research on LLM applications for Verilog code generation has expanded rapidly,
with major advancements emerging in the last two years.
This section outlines key methodologies and benchmarks in this evolving field,
with representative works summarized in \tabref{tab:methods}.

The methodologies are broadly classified into three approaches based on their application of LLMs: \emph{1) human interaction}, \emph{2) fine-tuned generation}, and \emph{3) iterative search}.
Although human interaction emerged with early adoption of LLM,
the latter two approaches have gained prominence more recently as researchers pursue greater automation.

\bheading{Human Interaction.}
Early efforts, such as ChipChat and ChipGPT~\cite{blocklove:2023:chipchat,chang:2023:chipgpt}, employed LLMs as copilots to assist engineers in design workflows.
These tools rely heavily on human expertise to complete designs through interactive prompts.

\bheading{Fine-tuned Generation.}
To further enhance the performance of general-purpose LLMs, researchers have explored fine-tuning of LLMs on hardware design datasets.
For example, \citet{liu:2023:chipnemo} introduced ChipNeMo, which fine-tunes a general-purpose LLM on internal NVIDIA datasets for various chip design tasks.
Similarly, \citet{thakur:2024:verigen} developed VeriGen to improve Verilog generation capabilities.
Subsequent works, such as RTLCoder~\cite{liu:2024:rtlcoder} and BetterV~\cite{pei:2024:betterv}, now represent the state-of-the-art in this category.

\bheading{Iterative Search.}
Iterative search approaches have recently gained prominence as a means to overcome the inherent limitations of one-shot generation.
AutoChip employs automated code refinement through error feedback~\cite{thakur:2023:autochip}.
RTLFixer employs a reasoning and acting framework to facilitate error correction~\cite{tsai:2024:rtlfixer}.

For benchmarking purposes, widely recognized datasets such as VerilogEval~\cite{liu:2023:verilogeval} and RTLLM~\cite{yao:2024:rtllm} are commonly utilized. These datasets have undergone significant evolution, with subsequent versions such as VerilogEvalV2~\cite{pinckney:2024:verilogevalv2} and RTLLM-2.0~\cite{liu:2024:openllm} being released. For the purpose of comparison, this study concentrates on the original versions of these datasets.

\subsection{Generalizing General-Purpose Code Generation}
Although not directly applicable to hardware design, research on general-purpose code generation provides valuable insights that can inform our work.
Population-based mechanisms, such as those utilized in FunSearch~\cite{romera:2024:mathematical} and EoH~\cite{liu:2024:eoh}, alongside tree-based reasoning frameworks like PlanSearch~\cite{wang:2024:planning} and ReAct prompting~\cite{yao:2023:react}, offer methodological inspiration.
These approaches collectively inspire our research direction.


\section{Evolution of Verilog (EvoVerilog)}

\subsection{Overview}
\begin{figure*}[t]
    \centering
    \includegraphics[width=0.96\textwidth]{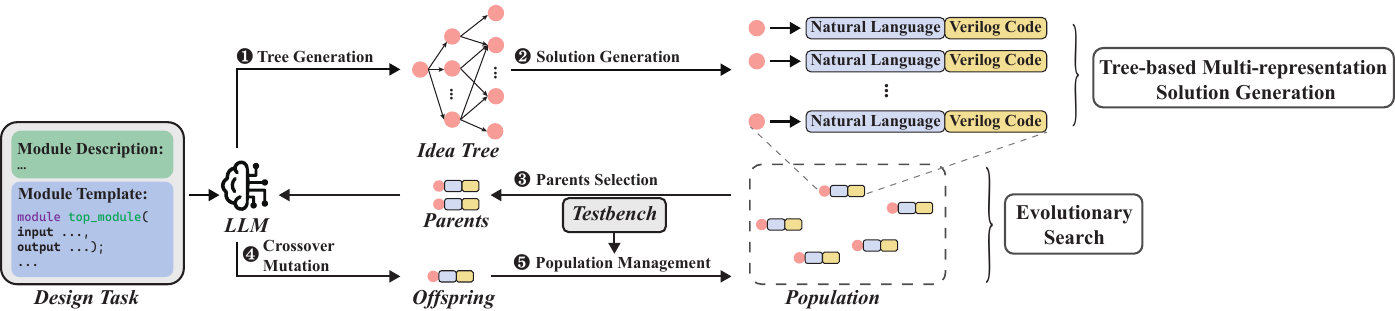}
    \caption{An illustrative overview of EvoVerilog. It initializes solutions through tree-based multi-representation generation, then iteratively refines them within the evolutionary search process.\label{fig:overview}}
\end{figure*}

Evolution of Verilog (EvoVerilog) is a framework that leverages the reasoning abilities of LLMs within the EA mechanism to iteratively generate Verilog code for hardware design.
The framework addresses two critical challenges in automated hardware design: \emph{exploring diverse design solutions} and \emph{optimizing for functional correctness and resource efficiency}.
Specifically, EvoVerilog operates in two phases: \emph{1)} tree-based multi-representation solution generation and \emph{2)} an evolutionary search mechanism.
An illustration of the end-to-end workflow is shown in \figref{fig:overview}.

\subsection{Tree-based Multi-Representation Generation}
Our tree-based solution generation process is designed to produce diverse and high-quality initial solutions for evolutionary search across multiple representation spaces.
Although directly using LLMs with task descriptions for solution generation is a straightforward approach,
it often results in solutions that are either overly similar or of low quality.
To address these limitations, we employ an idea tree to guide the search process across multiple representation spaces for the initialization of the solutions.

\subsubsection{Idea Tree}
While LLMs can generate numerous ideas for a given task, their output often suffer from two limitations: limited diversity of ideas and the absence of an efficient data structure to organize them effectively.
Our framework builds upon concepts from Tree-of-Thought reasoning~\cite{yao:2024:tree,wang:2024:planning},
with a key distinction: instead of relying entirely on systematic tree traversal, we employ the idea tree as a mechanism for probabilistically sampling high-quality initial solutions to seed evolutionary search. The ablation studies presented in \secref{subsec:ablation} demonstrate the efficacy of this approach, showing that EvoVerilog outperforms methods based solely on tree traversal.

\bheading{Idea Tree Generation.}
The tree is constructed through a recursive process, where each level refines or expands upon the ideas from the preceding levels.
The generation process unfolds as follows:
\begin{itemize}
    \item \textit{Level $1$}: Generate $N_0$ root ideas conditioned on the task description, establishing foundational concepts.
    \item \textit{Level $i$ $(i\geq 2)$}: Create variations by expanding at most the parent nodes $k$ from level $i-1$, resulting in potential combinations expressed as $\sum_{j=0}^k C_{N_{i-1}}^j$, where $N_{i-1}$ denotes the number of nodes at level $i-1$.
\end{itemize}

This structure exhibits exponential combinatorial growth with increasing depth,
rendering exhaustive search computationally impractical and leaving insufficient resources for solution refinement.
To mitigate this issue, we employ random sampling of $M$ solutions from the tree, which are subsequently translated into multiple representations for evolutionary search.

\subsubsection{Multi-representation Generation.}
The guiding ideas from the tree must be transformed into final implementations.
Although direct generation of Verilog code is possible,
code that lacks descriptive annotations or supplementary materials (\eg, logic diagrams) becomes challenging to interpret and modify.
Recent studies in the field of LLM research have demonstrated that the integration of natural language significantly improves the efficacy of code generation~\cite{wang:2024:planning,liu:2024:eoh,romera:2024:mathematical}.
We therefore introduce a dual-representation framework:

\begin{itemize}
    \item \emph{Indirect representation}: This includes natural language, logical diagrams, or mathematical formulas, which provide richer contextual information.
    \item \emph{Direct representation}: This consists of executable Verilog code for immediate evaluation.
\end{itemize}

\bheading{Indirect Solution Generation.}
Indirect code representations enhance contextual understanding, thereby improving both comprehension and adaptability.
This approach also facilitates the efficient exploration of the conceptual design space, as evidenced in \cite{wang:2024:planning}.
Notably, the historical limitations of evolutionary algorithms in Verilog optimization, as discussed by \citet{nielsen:2016:genetic}, arise from the structural similarity between the embedding space of Genetic Programming and the sparse search space of Verilog,
which hinders effective exploration.

While LLMs can generate code in various formats, we prioritize natural language due to its flexibility, expressiveness, and foundational role in the reasoning processes.

\bheading{Direct Solution Generation.}
In our framework, the direct representation comprises standardized Verilog code.
To ensure consistency across representations,
LLMs are explicitly instructed to align the generated code with both the underlying conceptual ideas and their corresponding natural language descriptions.

\subsection{Evolutionary Search}
Our evolutionary search framework methodically maintains and iteratively refines a population of diverse hardware design solutions through systematic optimization.
Although LLM-based evolutionary algorithms have been previously employed in general-purpose programming as demonstrated in prior studies~\cite{ye:2024:reevo,liu:2024:eoh,guo:2024:coevo},
our approach uniquely incorporates specialized offspring operators and selection mechanisms specifically designed for diverse hardware design contexts.
As illustrated in \algref{alg:es},
the process synergizes tree-based initialization with evolutionary operators to effectively balance exploration and exploitation in pursuit of optimal solutions.

\renewcommand{\algorithmiccomment}[1]{// #1}
\begin{algorithm}[t]
    \caption{Evolutionary Search of EvoVerilog}
    \label{alg:es}
    \begin{algorithmic}[1]
        \STATE {\bfseries Input:} A multi-representation solution set $P$ from the tree-based multi-representation generation
        \STATE {\bfseries Output:} Optimized solution set $P$
        \REPEAT
        \STATE $O_1 \leftarrow \textsc{Tree-based\_INIT}()$ \COMMENT{\textcolor{blue}{or $O_1 \leftarrow \emptyset$}}
        \STATE  $O_2\leftarrow \textsc{OffSpring}(P)$
        \STATE $O \leftarrow O_1 \cup O_2$
        \STATE $C \leftarrow P \cup O$
        \STATE $P \leftarrow \textsc{Non-dominated\_Selection}(C)$
        \UNTIL{Termination criteria met}
    \end{algorithmic}
\end{algorithm}

\subsubsection{Algorithm Overview}
Our evolutionary search algorithm operates through a defined sequence consisting of three iterative phases, each crucial for generating and refining solutions within a diverse design space:
\begin{enumerate}
    \item \textit{Solution Initialization} \textbf{(Optional)}: This phase seeds new solutions derived from the idea tree, initializing a foundation for further evolutionary processes.
    \item \textit{Offspring Generation}: In this phase, new solutions are produced through the application of crossover and mutation operators, expanding the solution space with novel variants.
    \item \textit{Solution Selection}: This final phase evaluates the solutions, selecting the most promising ones based on quality through a dominance-based filtering method.
\end{enumerate}

\subsubsection{Offspring Generation.}
The process of generating offspring involves two primary steps: \emph{(1)} parent selection and \emph{(2)} offspring creation. While traditional selection methods, \eg, tournament selection, are still relevant, this study focuses on the development of specialized mechanisms of generating offspring.
Our framework incorporates four distinct operators: \emph{(1)} positive crossover, \emph{(2)} negative crossover, \emph{(3)} positive mutation, and \emph{(4)} negative mutation.

\bheading{Crossove \& Mutation.}
These operators systematically recombine and modify existing solutions to generate novel design candidates.
Positive operators utilize the parents' design characteristics,
whereas negative operators intentionally diverge from parent solutions to promote exploration.
For detailed examples of these operator implementations, please refer to Appendix.~\ref{app:offspring}.

\subsubsection{Solution Selection.}
To preserve both diversity and quality within the population across iterations, a non-dominated sorting mechanism is employed for solution selection.
This method effectively balances two critical objectives: (1) minimizing the mismatch rate from the test bench and (2) minimizing hardware resource consumption, \eg, wires and cells in the Verilog module.
Traditional single-objective optimization approaches fail to adequately address the trade-offs between these competing objectives, thus making non-dominated sorting pivotal in identifying optimal Pareto solutions.

Before introducing non-dominated sorting, it is essential to define the concept of Pareto optimality for comparison.

\begin{definition}[Pareto Optimal]
    A solution $x^*$ is Pareto optimal if there exists no other solution $x$ such that $x_i \leq x^*_i$ for all $i\in\{1,\ldots, m\}$ and $x_j < x^*_j$ for at least one $j\in\{1,\ldots, m\}$.
\end{definition}

\begin{definition}[Pareto Set and Pareto Front]
    The Pareto set comprises all Pareto optimal solutions, and the Pareto front represents the objective function values of these solutions.
\end{definition}

\bheading{Non Dominated Sorting.}
The process begins by evaluating the dominance relationships between all pairs of solutions.
Solutions not dominated by any other form the first \textit{Pareto front} ($\mathcal{F}_1$).
Subsequent fronts are iteratively constructed by removing $\mathcal{F}_1$ and identifying the new nondominated set ($\mathcal{F}_2$), repeating until all solutions are classified.
This structure enables selective pressure toward the Pareto frontier and promotes diversity through front-based stratification.
Eventually, the best $N$ solutions across the fronts are preserved, prioritizing earlier fronts and employing random selection within each front to avoid premature convergence.

\begin{table*}[t]
    \centering
    \caption{Comparative analysis of Verilog code generation performance. Gray highlighting denotes state-of-the-art results (bold), underlines indicate second-best. Color-coded arrows show performance deltas relative to base models (green: improvement, red: decline).\label{tab:overall}}
    \vspace{0.25\baselineskip}
    \resizebox{0.98\textwidth}{!}{
        \begin{tabular}{llllcccccc}
            \toprule
            \multirow[c]{2}{*}{\textbf{Category}}              &                                            & \multirow[c]{2}{*}{\textbf{Method}} & \multirow[c]{2}{*}{\textbf{Params.}} & \multicolumn{3}{c}{\textbf{VerilogEval-Machine}}         & \multicolumn{3}{c}{\textbf{VerilogEval-Human}}                                                                                                                                                                                                                                                                                                                         \\
            \cmidrule{5-7}\cmidrule{8-10}
                                                               &                                            &                                     &                                      & \textbf{pass@1}                                          & \textbf{pass@5}                                                           & \textbf{pass@10}                                                          & \textbf{pass@1}                                          & \textbf{pass@5}                                                          & \textbf{pass@10}                                                         \\
            \midrule
            \multirow[c]{6}{*}{\textbf{Base Model}}            & \multirow[c]{3}{*}{\textbf{Closed-Source}} & GPT-3.5-Turbo                       & N/A                                  & 63.5                                                     & 78.0                                                                      & 81.6                                                                      & 31.2                                                     & 47.0                                                                     & 53.7                                                                     \\
                                                               &                                            & GPT-4o-mini                         & N/A                                  & 66.0                                                     & 72.4                                                                      & 74.0                                                                      & 54.2                                                     & 62.0                                                                     & 64.6                                                                     \\
                                                               &                                            & GPT-4-Turbo                         & N/A                                  & 72.5                                                     & \underline{83.0}                                                          & \underline{85.8}                                                          & 64.3                                                     & \underline{76.1}                                                         & \underline{79.7}                                                         \\
            \cmidrule{2-10}
                                                               & \multirow[c]{3}{*}{\textbf{Open-Source}}   & CodeQwen1.5-7B-Chat                 & 7B                                   & 29.1                                                     & 61.9                                                                      & 71.8                                                                      & 14.8                                                     & 36.8                                                                     & 45.0                                                                     \\
                                                               &                                            & DeepSeek-Coder                      & 6.7B                                 & 8.8                                                      & 34.3                                                                      & 53.8                                                                      & 4.9                                                      & 19.3                                                                     & 30.9                                                                     \\
                                                               &                                            & DeepSeek-V3                         & 671B                                 & \cellcolor{light-gray}\textbf{79.2}                      & 80.7                                                                      & 81.6                                                                      & \cellcolor{light-gray}\textbf{66.1}                      & 72.1                                                                     & 73.9                                                                     \\

            \midrule
            \multirow[c]{5}{*}{\textbf{Fine-tuned Generation}} & \multirow[c]{3}{*}{\textbf{Closed-Source}} & ChipNeMo$^\dagger$                  & 70B                                  & 53.8                                                     & N/A                                                                       & N/A                                                                       & 27.6                                                     & N/A                                                                      & N/A                                                                      \\
                                                               &                                            & VerilogEval$^\dagger$               & 16B                                  & 46.2                                                     & 67.3                                                                      & 73.7                                                                      & 28.8                                                     & 45.9                                                                     & 52.3                                                                     \\
                                                               &                                            & BetterV-CodeQwen$^\dagger$          & 7B                                   & $\text{68.1}_{\textcolor{Green}{\text{+39.0}}}$          & $\text{79.4}_{\textcolor{Green}{\text{+17.5}}}$                           & $\text{84.5}_{\textcolor{Green}{\text{+12.7}}}$                           & $\text{46.1}_{\textcolor{Green}{\text{+31.3}}}$          & $\text{53.7}_{\textcolor{Green}{\text{+16.9}}}$                          & $\text{58.2}_{\textcolor{Green}{\text{+13.2}}}$                          \\

            \cmidrule{2-10}
                                                               & \multirow[c]{2}{*}{\textbf{Open-Source}}   & VeriGen$^\dagger$                   & 16B                                  & 44.0                                                     & 52.6                                                                      & 59.2                                                                      & 30.3                                                     & 43.9                                                                     & 49.6                                                                     \\
                                                               &                                            & RTLCoder-DeepSeek-Coder             & 6.7B                                 & $\text{37.2}_{\textcolor{Green}{\text{+28.4}}}$          & $\text{64.9}_{\textcolor{Green}{\text{+30.6}}}$                           & $\text{74.8}_{\textcolor{Green}{\text{+21.0}}}$                           & $\text{16.9}_{\textcolor{Green}{\text{+12.0}}}$          & $\text{35.7}_{\textcolor{Green}{\text{+16.4}}}$                          & $\text{43.3}_{\textcolor{Green}{\text{+12.4}}}$                          \\
            \midrule
            \multirow[c]{5}{*}{\textbf{Iterative Search}}      & \multirow[c]{5}{*}{\textbf{Open-Source}}   & AutoChip-GPT-4o-mini                & N/A                                  & $\text{61.7}_{\textcolor{Red}{\text{-4.3}}}$             & $\text{69.0}_{\textcolor{Red}{\text{-3.4}}}$                              & $\text{70.6}_{\textcolor{Red}{\text{-3.4}}}$                              & $\text{49.4}_{\textcolor{Red}{\text{-4.8}}}$             & $\text{57.1}_{\textcolor{Red}{\text{-4.9}}}$                             & $\text{60.2}_{\textcolor{Red}{\text{-4.4}}}$                             \\
                                                               &                                            & AutoChip-DeepSeek-V3                & 671B                                 & \underline{$\text{77.8}_{\textcolor{Red}{\text{-1.4}}}$} & $\text{82.7}_{\textcolor{Green}{\text{+2.0}}}$                            & $\text{85.6}_{\textcolor{Green}{\text{+4.0}}}$                            & \underline{$\text{63.1}_{\textcolor{Red}{\text{-3.0}}}$} & $\text{70.3}_{\textcolor{Red}{\text{-1.8}}}$                             & $\text{74.6}_{\textcolor{Green}{\text{+0.7}}}$                           \\
            \cmidrule{3-10}
                                                               &                                            & \textbf{EvoVerilog-GPT-3.5-Turbo}   & N/A                                  & $\text{39.2}_{\textcolor{Red}{\text{-24.3}}}$            & $\text{74.1}_{\textcolor{Red}{\text{-4.4}}}$                              & $\text{80.3}_{\textcolor{Red}{\text{-1.3}}}$                              & $\text{21.1}_{\textcolor{Red}{\text{-10.1}}}$            & $\text{47.9}_{\textcolor{Green}{\text{+0.9}}}$                           & $\text{55.1}_{\textcolor{Green}{\text{+1.4}}}$                           \\
                                                               &                                            & \textbf{EvoVerilog-GPT-4o-mini}     & N/A                                  & $\text{45.6}_{\textcolor{Red}{\text{-20.4}}}$            & $\text{72.6}_{\textcolor{Green}{\text{+0.2}}}$                            & $\text{77.4}_{\textcolor{Green}{\text{+3.4}}}$                            & $\text{41.6}_{\textcolor{Red}{\text{-12.6}}}$            & $\text{61.7}_{\textcolor{Red}{\text{-0.3}}}$                             & $\text{65.4}_{\textcolor{Green}{\text{+0.8}}}$                           \\
                                                               &                                            & \textbf{EvoVerilog-DeepSeek-V3}     & 671B                                 & $\text{50.8}_{\textcolor{Red}{\text{-28.4}}}$            & \cellcolor{light-gray} $\textbf{84.2}_{\textcolor{Green}{\textbf{+3.5}}}$ & \cellcolor{light-gray} $\textbf{89.1}_{\textcolor{Green}{\textbf{+7.5}}}$ & $\text{52.0}_{\textcolor{Red}{\text{-14.1}}}$            & \cellcolor{light-gray}$\textbf{76.7}_{\textcolor{Green}{\textbf{+4.6}}}$ & \cellcolor{light-gray}$\textbf{80.2}_{\textcolor{Green}{\textbf{+6.3}}}$ \\

            \bottomrule
            \multicolumn{7}{l}{$^\dagger$: Reported Results.}                                                                                                                                                                                                                                                                                                                                                                                                                                                                                                                                                                \\
        \end{tabular}
    }
\end{table*}

\section{Experiment}
We conducted comprehensive experiments to evaluate EvoVerilog's effectiveness in generating functionally correct and resource-efficient Verilog code.
The experimental settings are detailed in \secref{subsec:exp_setting}.
The evaluation process consists of four phases:
First, we compare our method with key baseline approaches in different categories using the widely adopted pass@k metric, as described in \secref{subsec:overall_res}.
Second, we investigate the performance limits of EvoVerilog in Verilog generation through limited evaluation experiments on some challenging problems,
as outlined in \secref{subsec:unlimited_res}.
Third, we validate EvoVerilog's resource-efficiency capabilities by analyzing hardware resource utilization under unlimited evaluation settings, as detailed in \secref{subsec:multi_res}.
Finally, we performed ablation studies to examine the impact of the core components of EvoVerilog, as discussed in \secref{subsec:ablation}.

\subsection{Experimental Setting}\label{subsec:exp_setting}
\bheading{Dataset.}
For performance comparisons, we use VerilogEval~\cite{liu:2023:verilogeval}, which is composed of two distinct subsets: \textbf{machine}-generated and \textbf{human}-authored module descriptions. While the machine-generated subset contains detailed specifications, whereas the human-authored subset poses more significant challenges owing to its sparse manual contextual information.

\bheading{Baseline Methods.}
We compare our approach against three methodological categories: 1) standalone models, 2) fine-tuned generators, and 3) iterative search approaches. For closed source standalone models, we evaluated GPT-3.5-Turbo~\cite{gpt_3.5_turbo}, GPT-4o-mini~\cite{gpt_4o_mini}, and GPT-4-Turbo~\cite{gpt_4_turbo}. Open-source alternatives include CodeQwen1.5-7B-Chat~\cite{bai:2023:qwen}, DeepSeek-Coder~\cite{guo:2024:deepseek}, and DeepSeek-V3~\cite{liu:2024:deepseek}. The system prompt used for all models is the same, detailed in the Appendix~\ref{app:sys_prompt}. We also include a discussion of the influence of prompt design on the results in \secref{sec:discussion}.

For fine-tuned generation, we consider both closed-source methods (ChipNeMo~\cite{liu:2023:chipnemo}, VerilogEval~\cite{liu:2023:verilogeval}, Better~\cite{pei:2024:betterv}) and open-source implementations (VeriGen~\cite{thakur:2024:verigen}, RTLCoder~\cite{liu:2024:rtlcoder}). Our iterative search baseline combines AutoChip~\cite{thakur:2023:autochip} with the GPT-4o-mini and DeepSeek-V3 backbones.

\bheading{Hyper-parameters.}
For idea tree generation, we allow a maximum of $5$ ideas per node and up to $2$ parent nodes for derivations. Although this setting is consistent across all models, we adjusted the population size and offspring generation rate for evolutionary search within the allowed number of samples to ensure fair comparisons. The detailed hyperparameters are provided in the Appendix.~\ref{app:hyper}.

We access GPT models and DeepSeek-V3 via API, while reproducing open source implementations on an Ubuntu 22.04.6 LTS workstation (Intel Xeon E5-2697 v2, 124GB RAM, NVIDIA 3090Ti 24GB). Closed source results are reported from original publications.
Notably, VeriGen results are cited from the original paper due to hardware constraints.

\bheading{Evaluation Metrics.}
We adopt the widely-adopted pass@k metric in code generation tasks:
\begin{equation}
    \text{pass@k} = \frac{1}{N}\sum_{i=0}^N (1  -\frac{C_{n_i-c_i}^k}{C_{n_i}^k})
\end{equation}
where $N$ is the number of the problems in the dataset, $n_i$ and $c_i$ represent the total number of samples and correct samples for the $i$-th problem, respectively.

\subsection{Overall Results}\label{subsec:overall_res}
We evaluate EvoVerilog against state-of-the-art baselines using VerilogEval benchmark under a controlled sampling budget of 20 solutions per problem. Each method begins with 10 initial generations, with up to 20 total attempts for problems without valid solutions. The overall results are summarized in \tabref{tab:overall}.

\bheading{State-of-the-Art Performance.}
EvoVerilog establishes new benchmarks with DeepSeek-V3, achieving unprecedented pass@10 scores of 89.1 (machine subset) and 80.2 (human subset) - surpassing all existing approaches, including the 85.8/79.7 baseline from GPT-4 Turbo.
Notably, it outperforms AutoChip on these subsets despite using identical backbone models, demonstrating the effectiveness of evolutionary reasoning over simple iterative refinement.

\paragraph{Model-Scale Synergy} The framework exhibits a strong positive correlation with the capability of the model: While GPT-4o-mini shows moderate gains (+3.4 pass@10 on machine subset), DeepSeek-V3 achieves dramatic +7.5/+6.3 improvements.
This scaling law suggests evolutionary methods particularly benefit from strong base models' reasoning capabilities, unlike AutoChip which shows performance degradation (-3.4/-4.4 pass@10 for GPT-4o-mini).

\begin{figure}[t]
    \centering
    \includegraphics[width=0.46\textwidth]{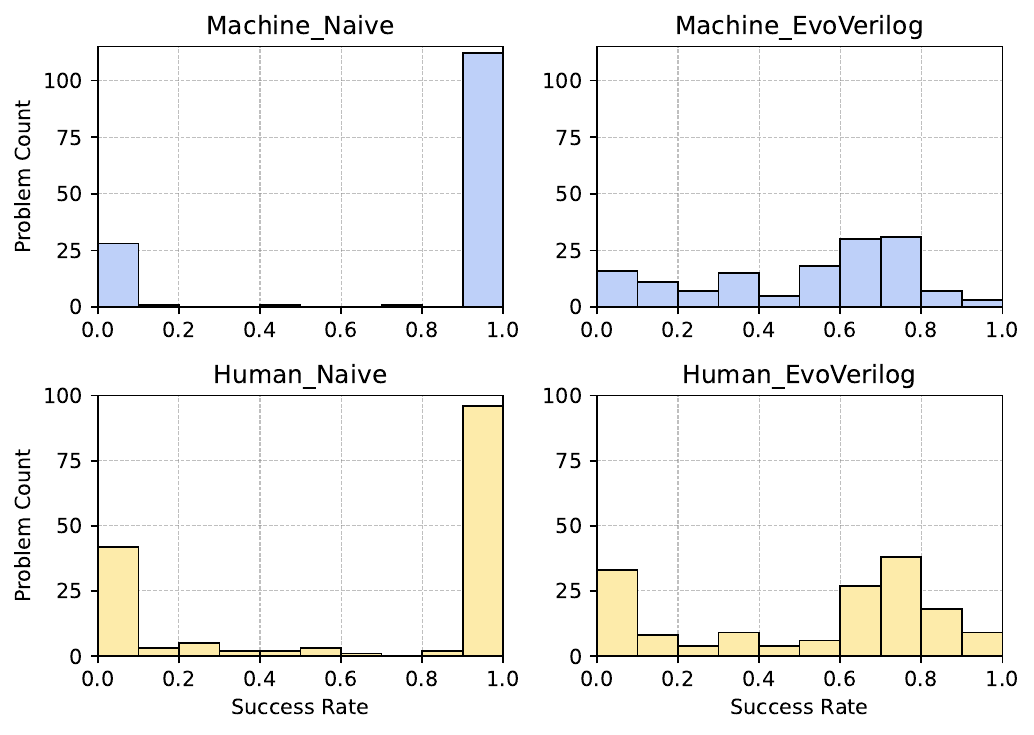}
    \caption{Success rate distribution analysis comparing EvoVerilog vs. naive sampling approaches using DeepSeek-V3 on two subsets.\label{fig:success_rate}}
\end{figure}

\paragraph{Diversity-Accuracy Tradeoff} A distinctive feature of EvoVerilog is observed in the pass@k trajectory: While there is a reduction of 28.4/14.1 points in pass@1 for DeepSeek-V3, pass@10 sees an increase of 7.5/6.3 points. This inverse relationship confirms our hypothesis that evolutionary optimization sacrifices some initial accuracy to discover diverse solutions, thus broadening the scope of correct implementations. Further support is provided by the success rate distributions (\figref{fig:success_rate}), where EvoVerilog shows significantly higher variance compared to the polarized outcomes of the baseline methods.

\paragraph{Practical Implications} The results challenge conventional assumptions about iterative refinement. AutoChip's negative performance deltas (-1.4 to -4.9 across metrics) suggest naively increasing sampling attempts for simple feedback refinement may degrade solution quality. In contrast, EvoVerilog's evolutionary approach provides reliable gains (+3.5 to +7.5 pass@10) without model retraining, making it immediately applicable to commercial LLM APIs. Our findings establish evolutionary reasoning as a new paradigm for hardware generation tasks.

\begin{table}
    \centering
    \caption{Extended evaluation results showing EvoVerilog's impact on total problem coverage. "Hard Solved" indicates hard problems addressed, "Final Success" shows final success rate including baseline solutions.\label{tab:resource_unlimited}}
    \vspace{0.25\baselineskip}
    \resizebox{0.48\textwidth}{!}{
        \begin{tabular}{lcccc}
            \toprule
            \multirow{2}{*}{\textbf{Backbone}} & \multicolumn{2}{c}{\textbf{Machine-Hard}} & \multicolumn{2}{c}{\textbf{Human-Hard}}                                                 \\
            \cmidrule{2-3}\cmidrule{4-5}
                                               & \textbf{Hard Solved}                      & \textbf{Final Success}                  & \textbf{Hard Solved} & \textbf{Final Success} \\
            \midrule
            GPT-3.5-Turbo                      & 38/55                                     & 88.11\%                                 & 53/102               & 68.59\%                \\
            GPT-4o-mini                        & 37/55                                     & 87.41\%                                 & 63/102               & 75.00\%                \\
            DeepSeek-V3                        & 47/55                                     & \textbf{94.41\%}                        & 73/102               & \textbf{81.41\%}       \\
            \bottomrule
        \end{tabular}
    }

\end{table}

\subsection{Extended Evaluation of Challenging Problems}\label{subsec:unlimited_res}
To assess EvoVerilog's upper performance boundaries, we conduct an extended evaluation with 100 solution attempts per problem and focus on a challenging benchmark - problems unsolved by the capable DeepSeek-Coder baseline (55 machine-subset and 102 human-subset problems, listed in Appendix.~\ref{app:challenging_problems}). The evaluation results are reported in \tabref{tab:resource_unlimited}.

\bheading{Results.}
EvoVerilog significantly improves the backbone model's ability to generate Verilog code. For instance, GPT-3.5-Turbo successfully solved 38 out of 55 machine-hard problems and 53 out of 102 human-hard problems, achieving final success rates of 88.11\% and 68.59\%, respectively, across the entire dataset.

Furthermore, the performance improves with more advanced models, as demonstrated by EvoVerilog-DeepSeek-V3, which solved 47 out of 55 machine-hard problems and 73 out of 102 human-hard problems. DeepSeek-V3 achieves absolute gains of 6.0–12.8\% over GPT-4o-mini across subsets, indicating that evolutionary methods more effectively leverage the reasoning capabilities of large models.

Notably, human-hard problems remain more challenging, with a success rate of 81.4\% compared to 94.4\% for machine-hard problems, suggesting that real-world designs require a deeper semantic understanding that current LLMs still struggle to achieve.

\begin{table}[t]
    \centering
    \caption{Resource efficiency performance on hard benchmarks. "Reduced" shows designs with resource savings, "Non-dominated" indicates Pareto-optimal solutions balancing wire/cell usage.\label{tab:reduced}}
    \vspace{0.25\baselineskip}
    \resizebox{0.48\textwidth}{!}{
        \begin{tabular}{lcccc}
            \toprule
            \multirow{2}{*}{\textbf{Backbone}} & \multicolumn{2}{c}{\textbf{Machine-Hard}} & \multicolumn{2}{c}{\textbf{Human-Hard}}                                             \\
                                               & \textbf{Reduced}                          & \textbf{Non-dominated}                  & \textbf{Reduced} & \textbf{Non-dominated} \\
            \midrule
            GPT-3.5-Turbo                      & 14/38                                     & 1                                       & 13/53            & 0                      \\
            GPT-4o-mini                        & 10/37                                     & 0                                       & 25/63            & 1                      \\
            DeepSeek-V3                        & 14/47                                     & 0                                       & 29/73            & 1                      \\
            \bottomrule
        \end{tabular}
    }
\end{table}

\subsection{Resource Efficiency Analysis}\label{subsec:multi_res}
To evaluate EvoVerilog's ability to produce hardware-efficient implementations, we analyze resource utilization patterns (wires and logic cells) across solutions for challenging benchmarks.
Focusing on the hard problem subset where design flexibility exists, we investigate two types of designs:
\emph{resource-reduced designs} and \emph{non-dominated designs}.
Resource-reduced designs use fewer resources than baseline implementations, while non-dominated designs are Pareto-optimal and cannot be improved in both resource dimensions without trade-offs.
The results are shown in \tabref{tab:reduced}. For full results containing all the original and optimized solutions, please refer to Appendix~\ref{app:full_res}.

\bheading{Results.}
For the challenging human subset, EvoVerilog-DeepSeek-V3 engineered the most resource-efficient solutions in 29 out of 73 cases (39.7\%), successfully balancing functional correctness with resource conservation.

Nevertheless, the prevalence of non-dominated solutions is low across both categories. This shortage may be attributable to the nature of most VerilogEval problems, which typically conform to a single optimal resource utilization solution. These non-dominated solutions are detailed in Appendix~\ref{app:non_dominated} for those interested. These outcomes underline the necessity for a specialized dataset that encompasses a wider array of design options in forthcoming studies.

\begin{table}[t]
    \centering
    \caption{Component ablation (DeepSeek-V3 backbone).\label{tab:ablation}}
    \vspace{0.2\baselineskip}
    \resizebox{0.48\textwidth}{!}{
        \begin{tabular}{lcccccc}
            \toprule
            \multirow[c]{2}{*}{\textbf{Configuration}} & \multicolumn{3}{c}{\textbf{VerilogEval-Machine}} & \multicolumn{3}{c}{\textbf{VerilogEval-Human}}                                                                                                                                                                                                    \\
            \cmidrule{2-4}\cmidrule{5-7}
                                                       & \textbf{pass@1}                                  & \textbf{pass@5}                                & \textbf{pass@10}                               & \textbf{pass@1}                               & \textbf{pass@5}                                & \textbf{pass@10}                               \\
            \midrule
            \textbf{EvoVerilog}                        & $\text{50.8}_{\textcolor{Red}{\text{-28.4}}}$    & $\text{84.2}_{\textcolor{Green}{\text{+3.5}}}$ & $\text{89.1}_{\textcolor{Green}{\text{+7.5}}}$ & $\text{52.0}_{\textcolor{Red}{\text{-14.1}}}$ & $\text{76.7}_{\textcolor{Green}{\text{+4.6}}}$ & $\text{80.2}_{\textcolor{Green}{\text{+6.3}}}$ \\
            \textbf{w/o  EA}                           & $\text{64.4}_{\textcolor{Red}{\text{-14.8}}}$    & $\text{81.2}_{\textcolor{Green}{\text{+0.5}}}$ & $\text{86.7}_{\textcolor{Green}{\text{+5.1}}}$ & $\text{60.8}_{\textcolor{Red}{\text{-5.3}}}$  & $\text{75.1}_{\textcolor{Green}{\text{+3.0}}}$ & $\text{79.1}_{\textcolor{Green}{\text{+5.2}}}$ \\
            \textbf{w/o Idea Tree}                     & $\text{65.6}_{\textcolor{Red}{\text{-13.6}}}$    & $\text{75.8}_{\textcolor{Red}{\text{-4.9}}}$   & $\text{83.3}_{\textcolor{Green}{\text{+1.7}}}$ & $\text{51.8}_{\textcolor{Red}{\text{-14.3}}}$ & $\text{75.3}_{\textcolor{Green}{\text{+3.2}}}$ & $\text{78.7}_{\textcolor{Green}{\text{+4.8}}}$ \\

            \bottomrule                                                                                                                                                                                                                                                                                                                                       \\
        \end{tabular}
    }
\end{table}
\subsection{Ablation Studies}\label{subsec:ablation}
To assess the impact of two key components in EvoVerilog, ablation studies were conducted using the VerilogEval benchmark, employing DeepSeek-V3 as the backbone model. The findings, presented in \tabref{tab:ablation}, delineate distinct performance trends across various configurations.

\bheading{Diversity-Accuracy Tradeoff}
The introduction of both components decreases generation diversity, as indicated by lower pass@1 scores in both subsets of the benchmark.

\bheading{Performance Improvement}
The Idea Tree boosts performance significantly in scenarios involving modules with comprehensive specifications. Conversely, for modules described with less detail, typically provided by humans, the two components exhibit parallel efficacy.

The ablation study confirms that both components collectively contribute to overall performance improvements.
\section{Discussion} \label{sec:discussion}
\subsection{Cost of Models and Effects}
\begin{figure}[t]
    \centering
    \includegraphics[width=0.48\textwidth]{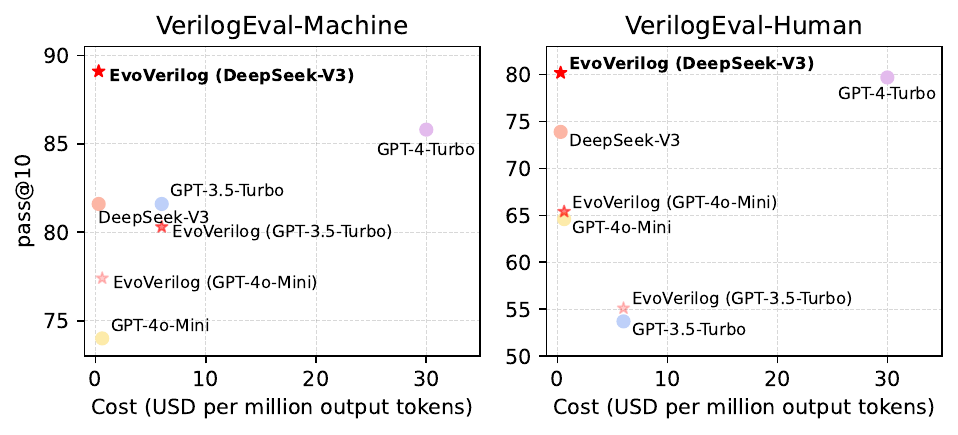}
    \caption{Cost-performance trade-off of the models. The plot compares API costs against pass@10 scores across baseline approaches and EvoVerilog.}
    \label{fig:cost}
\end{figure}

Understanding the cost implications of model selection is crucial for practical deployment.
To this end, we have detailed the costs of model APIs in Appendix.~\ref{app:cost}.
To demonstrate the cost-effectiveness and performance improvements facilitated by EvoVerilog, we have plotted both the cost and the pass@10 scores on the benchmark in \figref{fig:cost}.

Notably, DeepSeek-V3 emerges as a particularly cost-effective foundation model with robust reasoning capabilities,
which significantly enhances the performance of EvoVerilog.
The integration of EvoVerilog with DeepSeek-V3 as its backbone model has achieved state-of-the-art performance while maintaining a reasonable cost for practical deployment.

\subsection{Influence of Prompts}
The choice of prompts significantly influences the performance of Verilog code generation methods.
Since the prompts used by some methods in their original studies are not disclosed,
we reproduced their results using our own prompts.
To demonstrate the impact of prompt variation, we present the results of Standalone models using two distinct prompts in \tabref{tab:prompts}, with detailed descriptions of the testing prompts provided in the Appendix.~\ref{app:sys_prompt}.

The results confirm that the choice of prompts significantly impacts the performance of Verilog code generation.
This observation partially explains the discrepancies between our reproduced results and those reported in the original studies, which can be attributed to differences in prompt design.

Additionally, testing with diverse prompts for fine-tuned models provides valuable insights. A key takeaway from this work is the potential to enhance performance by integrating automated prompt optimization techniques.

\begin{table}[t]
    \centering
    \caption{Impact of prompt design on Verilog generation performance (pass@10 score).\label{tab:prompts}}
    \vspace{0.2\baselineskip}
    \resizebox{0.46\textwidth}{!}{
        \begin{tabular}{lcccc}
            \toprule
            \multirow{2}{*}{\textbf{Model}} & \multicolumn{2}{c}{\textbf{VerilogEval-Machine}} & \multicolumn{2}{c}{\textbf{VerilogEval-Human}}                             \\
            \cmidrule{2-3}\cmidrule{4-5}
                                            & \textbf{P1}                                      & \textbf{P0}                                    & \textbf{P1} & \textbf{P0} \\
            \midrule
            GPT-3.5-Turbo                   & 66.5                                             & 81.6                                           & 46.5        & 53.7        \\
            GPT-4o-mini                     & 64.1                                             & 74.0                                           & 51.8        & 64.6        \\
            GPT-4-Turbo                     & 70.1                                             & 85.8                                           & 69.1        & 79.7        \\
            DeepSeek-V3                     & 68.5                                             & 81.6                                           & 58.6        & 73.9        \\
            \bottomrule
        \end{tabular}
    }
\end{table}




\section{Conclusion}
We present EvoVerilog,
a framework integrating LLMs with evolutionary algorithms to automate Verilog code generation.
By combining tree-based exploration for diversity and non-dominated sorting for multi-objective optimization, EvoVerilog addresses the limitations of prior methods in design space exploration and resource efficiency.
On the VerilogEval benchmark, EvoVerilog achieves state-of-the-art pass@10 scores of 89.1 (machine) and 80.2 (human),
demonstrating its ability to generate both correct and resource-efficient solutions without human intervention.
This work advances hardware design automation, reducing manual effort while balancing functional and resource constraints,
and establishes a foundation for future research in LLM-driven EDA tools.

\section*{Impact Statement}

This paper presents work whose goal is to advance the field of Electronic Design Automation. There are many potential societal consequences of our work, none which we feel must be specifically highlighted here.

\bibliography{main}
\bibliographystyle{icml2025}

\newpage
\appendix
\onecolumn

\section{Prompts}

\subsection{Offspring Prompts} \label{app:offspring}
In this section we outlined the detailed prompts for the offspring operators.
Note that this generation process follows the ReAct framework~\cite{yao:2023:react}.
The offspring prompts are divided into two parts: reasoning and acting.
The reasoning prompts are designed to encourage the offspring to generate innovative and non-obvious observations about the problem.
Then, the acting prompts are designed to implement them in both indirect and direct representations.

The OFFSPRING\_PROMPT is a placeholder for different prompts for each offspring operator, which are detailed below:
\begin{itemize}
    \item \textbf{Positive Crossover:} creating a new solution that has a totally different form from the given solutions but can be motivated from the existing ones.
    \item \textbf{Negative Crossover:} creating a new solution that has a totally different form from the given solutions.
    \item \textbf{Positive Mutation:} creating a solution in different forms but can be a modified version of the existing solution.
    \item \textbf{Negative Mutation:} creating a totally different solution from the existing solution.
\end{itemize}

\begin{dialogbox}
    \textbf{Prompt for Offspring (Reasoning)} \\

    \textcolor{gray}{$\langle$SYS\_PROMPT$\rangle$}\\

    \textcolor{gray}{$\langle$TASK\_DESCRIPTION$\rangle$}\\

    \textcolor{gray}{$\langle$PARENT\_SOL\_0$\rangle$}\\
    \textcolor{gray}{$\langle$PARENT\_SOL\_1$\rangle$}\\
    \textcolor{gray}{...}\\

    \textcolor{red}{Brainstorm a list of \{num\_observations\} innovative and non-obvious observations about} \\

    \textcolor{gray}{$\langle$OFFSPRING\_PROMPT$\rangle$}
\end{dialogbox}

\begin{dialogbox}
    \textbf{Prompt for Offspring (Acting, indirect)} \\

    \textcolor{red}{Now, use the observations above to brainstorm a natural language solution to the problem above.}
\end{dialogbox}

\begin{dialogbox}
    \textbf{Prompt for Offspring (Acting, direct)} \\

    \textcolor{red}{Now, ONLY return the Verilog Code inside markdown codeblocks.}
\end{dialogbox}

\subsection{System Prompts} \label{app:sys_prompt}
To ensure a fair comparison, we use identical system prompts for all methods. However, some open-source methods do not provide system prompts. Therefore, we provide the system prompt \textbf{P0} for the baseline models in this section. Additionally, we manually created another system prompt, \textbf{P1}, to facilitate comparative analysis.

\begin{dialogbox}
    \textbf{System Prompt (Final adopted, P0)} \\
    You are an expert Verilog engineer. You will be given a problem, its details.  Avoid syntax errors and ensure the code is runnable. You will return Verilog code and passes all tests. \\

    Here is the problem:\\

    Module Description:\\

    \textcolor{gray}{$\langle$Description from VerilogEval$\rangle$}\\

    Module Template:\\

    \textcolor{gray}{$\langle$Template from VerilogEval$\rangle$}\\

    Return ONLY Verilog code inside markdown codeblocks.
\end{dialogbox}

\begin{dialogbox}
    \textbf{System Prompt (Comparative Analysis, P1)} \\

    You are an experienced Verilog engineer, please implement the 'top\_module' module based on the following description. Avoid syntax errors and ensure the code is runnable.\\

    Module Description:\\

    \textcolor{gray}{$\langle$Description from VerilogEval$\rangle$}\\

    Module Template:\\

    \textcolor{gray}{$\langle$Template from VerilogEval$\rangle$}\\

    Only respond the runnable code.\\

    Respond Example:\\

    \verb|```|verilog\\
    \textcolor{gray}{$\langle$Template from VerilogEval$\rangle$}\\
    ...\\
    \verb|```|
\end{dialogbox}

\section{Hyper-parameter settings}\label{app:hyper}
The following hyper-parameter configurations were employed under distinct generation limits:
\begin{itemize}
    \item \textbf{20 Samples:} For solutions derived from the idea tree, $M=6$. In generation 0, $6$ solutions were initialized from the idea tree. In generations 1 and 2, $3$ solutions were generated from the idea tree, and $4$ offspring were reproduced using the designed operators. This configuration yielded a total of $20$ samples.
    \item \textbf{100 Samples:} For solutions derived from the idea tree, $M=6$. In generation 0, $9$ solutions were initialized from the idea tree. In generations 1 through 13, $3$ solutions were generated from the idea tree, and $4$ offspring were reproduced using the designed operators. This configuration yielded a total of $100$ samples.
\end{itemize}

\section{Challenging problems}\label{app:challenging_problems}
We refer the problems that are unsolved by DeepSeek-Coder model within $20$ generations as challenging problems. The following table lists the challenging problems for both the machine-hard and human-hard subsets.

\begin{table}[h]
    \centering
    \caption{Challenging problems for VerilogEval benchmark.}
    \begin{tabular}{lp{13cm}}
        \textbf{Subset Name} & \textbf{Problems}                                                                                                                                                                                                                                                                                                                                                                                                                                                                                                                                                                                                                                                                                                                                                                                                                                                                                                                                                                                                                                                                                                                                                                                                                                                                                                                                                                                                                                                            \\
        \hline
        machine-hard         & 'mux2to1v', 'circuit7', 'ece241\_2014\_q5a', 'fsm3', 'ece241\_2013\_q8', 'm2014\_q6', 'fsm\_ps2data', 'kmap4', 'norgate', 'm2014\_q6c', '2014\_q4a', 'ece241\_2014\_q4', 'rule110', 'fsm3s', 'review2015\_fsmonehot', 'always\_casez', 'history\_shift', 'ece241\_2013\_q12', 'thermostat', 'ece241\_2013\_q2', 'lfsr32', 'gatesv', 'mt2015\_q4', 'rotate100', 'notgate', 'fsm\_onehot', 'popcount3', 'lemmings1', '7458', 'review2015\_fsmseq', 'count1to10', 'lfsr5', 'kmap3', 'wire', 'review2015\_fsmshift', '2013\_q2afsm', '2014\_q3bfsm', '2013\_q2bfsm', 'fsm3comb', 'dff16e', 'circuit8', 'gatesv100', 'm2014\_q4a', 'vectorgates', 'fsm3onehot', '2014\_q3fsm', 'vector4', 'always\_case2', 'fsm2', 'm2014\_q4d', 'shift4', 'fsm2s', 'vector3', 'ece241\_2014\_q5b', 'circuit10'                                                                                                                                                                                                                                                                                                                                                                                                                                                                                                                                                                                                                                                                                   \\
        human-hard           & 'gatesv', 'rotate100', 'review2015\_fsmonehot', 'dff8ar', 'kmap3', 'lemmings1', '2013\_q2afsm', 'fsm\_hdlc', 'review2015\_count1k', 'circuit8', '2014\_q3fsm', 'always\_if2', 'counter\_2bc', 'circuit1', '2012\_q1g', 'count1to10', 'm2014\_q4d', 'fsm1', 'dff16e', 'fsm\_onehot', 'review2015\_fancytimer', 'ece241\_2013\_q4', 'hadd', 'circuit9', 'ringer', 'fsm\_serial', 'timer', 'circuit7', 'gatesv100', 'vector100r', 'countslow', 'circuit3', 'mt2015\_q4', 'mt2015\_muxdff', 'fsm3comb', 'fsm\_serialdata', 'vector3', 'history\_shift', 'truthtable1', 'circuit5', 'm2014\_q3', 'm2014\_q4f', 'gshare', 'vector5', 'dff8p', 'circuit6', 'lemmings4', 'review2015\_fsm', 'fsm2', 'conwaylife', 'mt2015\_q4b', 'm2014\_q6', 'kmap2', 'ece241\_2014\_q4', 'rule110', 'review2015\_shiftcount', 'dff8', 'lemmings3', 'review2015\_fsmseq', 'fsm3', 'lfsr5', 'lfsr32', '2014\_q3bfsm', 'fsm2s', 'm2014\_q4b', 'm2014\_q4k', 'review2015\_fsmshift', 'count10', 'kmap4', 'wire', 'ece241\_2014\_q5b', 'ece241\_2014\_q3', 'circuit10', 'fsm\_ps2data', 'edgedetect', 'circuit4', 'fsm\_ps2', 'fsm3s', 'fsm3onehot', 'm2014\_q4a', 'edgecapture', 'vectorr', '2013\_q2bfsm', 'ece241\_2014\_q1c', 'ece241\_2013\_q2', 'm2014\_q6c', '2012\_q2fsm', 'count\_clock', 'rule90', 'lemmings2', 'ece241\_2014\_q5a', 'm2014\_q6b', 'circuit2', 'always\_if', 'ece241\_2013\_q12', 'shift18', 'ece241\_2013\_q8', 'fsm1s', 'countbcd', '2014\_q3c', 'thermostat', '2012\_q2b', \\
        \hline
    \end{tabular}
\end{table}

\section{Detailed Resource Efficiency Results}

\subsection{Full Results}\label{app:full_res}
We have included the baseline resource used by the original solutions and the optimized solutions generated by EvoVerilog-DeepSeek-V3 for both machine-hard and human-hard subsets of the VerilogEval benchmark.
\begin{table}[h]
    \centering
    \caption{Resource usage of original and optimized solutison. (DeepSeek-V3, Machine)}
    \vspace{0.2\baselineskip}
    \begin{tabular}{lcccc}
        \toprule
        \textbf{Problem ID}  & \multicolumn{2}{c}{\textbf{Original}} & \multicolumn{2}{c}{\textbf{Optimized}}                                 \\
        \cmidrule{2-3}\cmidrule{4-5}
                             & \textbf{Wire}                         & \textbf{Cell}                          & \textbf{Wire} & \textbf{Cell} \\
        \midrule
        mux2to1v             & 54                                    & 10                                     & 4             & 1             \\
        circuit7             & 5                                     & 2                                      & 4             & 2             \\
        ece241\_2014\_q5a    & 19                                    & 16                                     & 11            & 8             \\
        ece241\_2013\_q12    & 16                                    & 10                                     & 8             & 2             \\
        fsm\_onehot          & 16                                    & 23                                     & 9             & 16            \\
        popcount3            & 7                                     & 7                                      & 6             & 5             \\
        count1to10           & 6                                     & 4                                      & 5             & 3             \\
        kmap3                & 17                                    & 12                                     & 5             & 2             \\
        review2015\_fsmshift & 13                                    & 11                                     & 6             & 4             \\
        2014\_q3bfsm         & 20                                    & 17                                     & 16            & 13            \\
        dff16e               & 15                                    & 12                                     & 5             & 2             \\
        gatesv100            & 4                                     & 4                                      & 4             & 3             \\
        vectorgates          & 7                                     & 6                                      & 7             & 5             \\
        always\_case2        & 8                                     & 7                                      & 2             & 2             \\
        \bottomrule
    \end{tabular}
\end{table}

\begin{table}[h]
    \centering
    \caption{Resource usage of original and optimized solutison. (DeepSeek-V3, Human)}
    \vspace{0.2\baselineskip}
    \begin{tabular}{lcccc}
        \toprule
        \textbf{Problem ID}    & \multicolumn{2}{c}{\textbf{Original}} & \multicolumn{2}{c}{\textbf{Optimized}}                                 \\
        \cmidrule{2-3}\cmidrule{4-5}
                               & \textbf{Wire}                         & \textbf{Cell}                          & \textbf{Wire} & \textbf{Cell} \\
        \midrule
        gatesv                 & 4                                     & 10                                     & 4             & 3             \\
        circuit1               & 5                                     & 2                                      & 3             & 1             \\
        count1to10             & 13                                    & 11                                     & 5             & 3             \\
        fsm1                   & 10                                    & 6                                      & 9             & 4             \\
        dff16e                 & 15                                    & 12                                     & 5             & 2             \\
        hadd                   & 11                                    & 8                                      & 4             & 2             \\
        circuit9               & 12                                    & 12                                     & 7             & 6             \\
        timer                  & 15                                    & 12                                     & 10            & 7             \\
        circuit7               & 6                                     & 4                                      & 4             & 2             \\
        vector100r             & 12                                    & 0                                      & 2             & 0             \\
        mt2015\_q4             & 11                                    & 7                                      & 4             & 2             \\
        fsm3comb               & 13                                    & 10                                     & 4             & 3             \\
        vector3                & 12                                    & 0                                      & 11            & 0             \\
        truthtable1            & 10                                    & 7                                      & 7             & 4             \\
        circuit5               & 11                                    & 5                                      & 10            & 5             \\
        vector5                & 20                                    & 20                                     & 16            & 14            \\
        review2015\_shiftcount & 11                                    & 7                                      & 9             & 5             \\
        2014\_q3bfsm           & 17                                    & 14                                     & 16            & 13            \\
        m2014\_q4k             & 11                                    & 7                                      & 5             & 1             \\
        review2015\_fsmshift   & 11                                    & 9                                      & 10            & 7             \\
        count10                & 7                                     & 4                                      & 5             & 3             \\
        ece241\_2014\_q5b      & 18                                    & 15                                     & 11            & 8             \\
        ece241\_2014\_q1c      & 14                                    & 9                                      & 6             & 4             \\
        m2014\_q6c             & 11                                    & 9                                      & 8             & 6             \\
        2012\_q2fsm            & 18                                    & 15                                     & 18            & 14            \\
        fsm1s                  & 9                                     & 5                                      & 7             & 3             \\
        countbcd               & 20                                    & 20                                     & 15            & 18            \\
        2014\_q3c              & 15                                    & 12                                     & 8             & 5             \\
        2012\_q2b              & 11                                    & 9                                      & 8             & 6             \\
        \bottomrule
    \end{tabular}
\end{table}

\subsection{Non-dominated designs}\label{app:non_dominated}

\subsubsection{Problem: always\_case2, found by GPT-3.5-Turbo}

\begin{lstlisting}[language=Verilog, caption={always\_case2 (wire:7, cell:10)}]
    module top_module (
    input [3:0] in,
    output reg [1:0] pos
);

always @(*) begin
    case (in)
        2, 6, 10, 14:
            pos = 1;
        4, 12:
            pos = 2;
        8:
            pos = 3;
        default:
            pos = 0;
    endcase
end

endmodule
\end{lstlisting}

\begin{lstlisting}[language=Verilog, caption={always\_case2 (wire:2, cell:17)}]
    module top_module (
        input [3:0] in,
        output reg [1:0] pos
    );
    
    reg [1:0] pos_table [15:0]; // Lookup table for pos values
    
    initial begin
        // Initialize lookup table based on the given rules
        pos_table[0] = 2'b00;  pos_table[1] = 2'b00;
        pos_table[2] = 2'b01;  pos_table[3] = 2'b00;
        pos_table[4] = 2'b10;  pos_table[5] = 2'b00;
        pos_table[6] = 2'b01;  pos_table[7] = 2'b00;
        pos_table[8] = 2'b11;  pos_table[9] = 2'b00;
        pos_table[10] = 2'b01; pos_table[11] = 2'b00;
        pos_table[12] = 2'b10; pos_table[13] = 2'b00;
        pos_table[14] = 2'b01; pos_table[15] = 2'b00;
    end
    
    always @(*) begin
        pos = pos_table[in];
    end
    
    endmodule
\end{lstlisting}

\subsubsection{Problem: vector5, found by DeepSeek-V3}
\begin{lstlisting}[language=Verilog, caption={vector5 (wire:16, cell:20)}]
    module top_module (
        input a,
        input b,
        input c,
        input d,
        input e,
        output [24:0] out
    );
    
        // Generate all 25 pairwise comparisons
        assign out[24] = ~(a ^ a);
        assign out[23] = ~(a ^ b);
        assign out[22] = ~(a ^ c);
        assign out[21] = ~(a ^ d);
        assign out[20] = ~(a ^ e);
    
        assign out[19] = ~(b ^ a);
        assign out[18] = ~(b ^ b);
        assign out[17] = ~(b ^ c);
        assign out[16] = ~(b ^ d);
        assign out[15] = ~(b ^ e);
    
        assign out[14] = ~(c ^ a);
        assign out[13] = ~(c ^ b);
        assign out[12] = ~(c ^ c);
        assign out[11] = ~(c ^ d);
        assign out[10] = ~(c ^ e);
    
        assign out[9]  = ~(d ^ a);
        assign out[8]  = ~(d ^ b);
        assign out[7]  = ~(d ^ c);
        assign out[6]  = ~(d ^ d);
        assign out[5]  = ~(d ^ e);
    
        assign out[4]  = ~(e ^ a);
        assign out[3]  = ~(e ^ b);
        assign out[2]  = ~(e ^ c);
        assign out[1]  = ~(e ^ d);
        assign out[0]  = ~(e ^ e);
    
    endmodule
\end{lstlisting}

\begin{lstlisting}[language=Verilog, caption={vector5 (wire:20, cell:14)}]
    module top_module (
        input a,
        input b,
        input c,
        input d,
        input e,
        output [24:0] out
    );
    
        // Compute the unique comparisons
        wire ab = ~a ^ b;
        wire ac = ~a ^ c;
        wire ad = ~a ^ d;
        wire ae = ~a ^ e;
        wire bc = ~b ^ c;
        wire bd = ~b ^ d;
        wire be = ~b ^ e;
        wire cd = ~c ^ d;
        wire ce = ~c ^ e;
        wire de = ~d ^ e;
    
        // Assign the unique comparisons to the output vector
        assign out[24] = 1'b1; // a == a
        assign out[23] = ab;   // a == b
        assign out[22] = ac;   // a == c
        assign out[21] = ad;   // a == d
        assign out[20] = ae;   // a == e
        assign out[19] = ab;   // b == a
        assign out[18] = 1'b1; // b == b
        assign out[17] = bc;   // b == c
        assign out[16] = bd;   // b == d
        assign out[15] = be;   // b == e
        assign out[14] = ac;   // c == a
        assign out[13] = bc;   // c == b
        assign out[12] = 1'b1; // c == c
        assign out[11] = cd;   // c == d
        assign out[10] = ce;   // c == e
        assign out[9]  = ad;   // d == a
        assign out[8]  = bd;   // d == b
        assign out[7]  = cd;   // d == c
        assign out[6]  = 1'b1; // d == d
        assign out[5]  = de;   // d == e
        assign out[4]  = ae;   // e == a
        assign out[3]  = be;   // e == b
        assign out[2]  = ce;   // e == c
        assign out[1]  = de;   // e == d
        assign out[0]  = 1'b1; // e == e
    
    endmodule
\end{lstlisting}

\section{Cost of the Tokens of Standalone Models} \label{app:cost}
We provide the cost of tokens for standalone models in the following table.
\begin{table}[h]
    \caption{Cost of the tokens for the standalone models. The unit is dollars per million tokens.}
    \vspace{0.2\baselineskip}
    \centering
    \begin{tabular}{ll}
        \toprule
        \textbf{Standalone Model} & \textbf{Cost} \\
        \midrule
        GPT-3.5-Turbo             & 6.0           \\
        GPT-4o-mini               & 0.6           \\
        GPT-4-Turbo               & 30.0          \\
        DeepSeek-V3               & 0.28          \\
        \bottomrule
    \end{tabular}
\end{table}

\end{document}